\documentclass[letterpaper,prb,twocolumn,showpacs,amsmath,amssymb]{revtex4}

\usepackage{graphicx}
\usepackage{dcolumn}
\usepackage{bm}

\newcommand{\be}{\begin{equation}}
\newcommand{\ee}{\end{equation}}
\newcommand{\bea}{\begin{eqnarray}}
\newcommand{\eea}{\end{eqnarray}}
\newcommand{\mbf}{\mathbf}
\newcommand{\bs}{\boldsymbol}

\begin{document}

\title{
Local density of states of a strongly type-II $d$-wave superconductor: \\
The binary alloy model in a magnetic field
}
\author{J. Lages}
\thanks{Address after October 1st 2003:
\texttt{lages@ameslab.gov}, Ames Laboratory, Department of Physics
and Astronomy, Iowa State University, Ames IA 50011, USA.}
\author{P. D. Sacramento}
\email{pdss@cfif.ist.utl.pt} \affiliation{Centro de F\'\i sica das
Interac\c c\~oes Fundamentais, Instituto Superior T\'ecnico,
Avenida Rovisco Pais, 1049-001 Lisboa, Portugal}

\date{\today}

\begin{abstract}
We calculate self-consistently the local density of states (LDOS) of a $d$-wave superconductor
considering the scattering of the quasiparticles off randomly distributed
impurities and off externally induced vortices.
The impurities and the vortices
are randomly distributed but the vortices are preferably located near the impurities.
The increase of either the impurity repulsive potential or the impurity density
only affects the density of states (DOS) slightly.
The dominant effect is due to the vortex scattering.
The results for the LDOS agree qualitatively with experimental results considering that most
vortices are pinned at the impurities.
\end{abstract}


\maketitle

Experimental evidence
suggests that the pairing symmetry in high temperature superconductors (HTSC)
is $d$-wave \cite{dwave}. A good description of the non-conventional superconducting phase
is obtained using a standard BCS approach but a clear understanding of the normal phase
of these materials remains a hard challenge.
If we apply a strong enough magnetic field, these materials (being strongly
type-II superconductors) will enter a vortex phase.
Scanning Tunneling Microscopy (STM) studies of HTSC have revealed a very different quasiparticle
structure from that predicted by the pure $d$-wave BCS single
vortex models \cite{Hoffman1}. In a pure $d$-wave there are
extended gapless states, a fourfold symmetric shape of the LDOS
and a zero-bias conductance peak. Experimentally, however, it is
observed: i) absence of zero energy peaks, ii) absence of
coherence peaks close to the vortex, iii) low energy states with
an energy $\sim 5.5\textrm{meV}$ for YBCO and $\sim
7.7\textrm{meV}$ for BSCCO, and iv) absence of $4$-fold symmetric
star-shaped LDOS \cite{Hoffman1}. The coherence peaks are
recovered about $10\mathrm{\AA}$ from the core center \cite{Pan}
and the core states are localized decaying exponentially with
distance ($\sim 22\mathrm{\AA}$) \cite{Pan}. Possible reasons
for the failure of the pure $d$-wave theory to explain the
experimental results have included a mixed pairing of the type
$d_{x^2-y^2} + i d_{xy}$, considering that the
vortices have antiferromagnetic cores such that localized magnetic
order coexists with superconductivity or charge
order fluctuations \cite{Franz1,magcdw}.

Recently, a pure $d$-wave pairing for a vortex lattice has yielded
results that are in good qualitative agreement with experiments
\cite{Lages2}. Indeed a calculation of the local density of states
shows that close to the vortex position the coherence peaks disappear,
there are significant low energy peaks and there is no zero energy enhancement
of the density of states \cite{Lages2}. However,
in most systems disorder is present for instance in the form of
impurities. The presence of the impurities affects the quasiparticles (QP) in two
ways: the QP scatter off the impurities due to potential
scattering (if they are nonmagnetic) and the impurities pin the
vortices also affecting the density of states of the QP, particularly
at low energies.

The separate effects of the scattering of the quasiparticles from
the impurities and from the vortices have been studied before. The
details of the impurity disorder are relevant and a consistent
picture of the various possible scenarios has been obtained
\cite{Altland}.
Studies of the superfluid stiffness due to the presence of the
impurities have revealed that, even though it gets lower, the
decrease is smaller than expected. The reason is
that the order parameter is only significantly affected very close
to the impurities and largely unaffected elsewhere. Therefore the
order parameter is very non-homogeneous and a fully
self-consistent calculation is required \cite{Atkinson2}.
Considering exclusively the effect of the scattering off
homogeneously distributed vortices
(no impurities) it has been shown that in the lattice case the low
energy states are extended Bloch states \cite{Bloch} instead of
Dirac Landau levels \cite{Gorkov}. Also, it was shown that the
quasiparticles besides feeling a Doppler shift caused by the
moving supercurrents \cite{Volovik}, also feel a quantum "Berry"
like term due to a half-flux, $\phi_0/2$, Aharonov-Bohm scattering of the
quasiparticles by the vortices. The effect of a random vortex
distribution was considered
taking random and statistically independent
scalar and vector potentials \cite{K.}. A finite
density of states was predicted at zero energy. Also, considering
randomly pinned discrete vortices the density of states was
calculated displaying low energy peaks and no
coherence peaks \cite{aprile}
but at low energies a power law deviation from a
finite zero energy value was found, where both the zero energy
value and the exponent depend on the magnetic field and on the
Dirac anisotropy \cite{Lages2}.
Also, it was found that
even though the low energy states are strongly peaked at the
vortex cores, they appear to remain extended. An approximate
scaling of the density of states was found at low energies
\cite{Lages2}.

In this work we study the {\it combined} effects on the quasiparticle spectrum of the scattering
off impurities and vortices induced by an external magnetic field.
We model the disorder using the
binary alloy model \cite{Atkinson} where the impurities are distributed randomly
over the lattice sites. At each impurity site it costs an energy $U$ to place an
electron (it acts as a local shift on the chemical potential).
The impurities are randomly distributed over a $L\times L$ periodic
two-dimensional lattice and play the role of pinning centers for
the vortices.
It is favorable that a vortex
is located in the vicinity of an impurity \cite{pinning}.
Taking into account a given distribution of the
positions of the impurities
$\left\{\mathbf{r}^p_i\right\}_{i=1,N_p}$, the distribution of the
positions of the vortices
$\left\{\mathbf{r}^v_i\right\}_{i=1,N_v}$ is chosen in such a way
it minimizes the total vortex energy given by
$\mathcal{E}=\mathcal{E}_v+\mathcal{E}_p$,
where $\mathcal{E}_v=\mathcal{U}_v\sum_{\mathbf{r}^v_i,\mathbf{r}^v_j}K_0
\left(\left|\mathbf{r}^v_i-\mathbf{r}^v_j\right|/\lambda\right)$ is the repulsive
interaction energy between the vortices in the London regime
and $\mathcal{E}_p=\mathcal{U}_p\sum_{\mathbf{r}^v_i,\mathbf{r}^p_j}
    f\left(\left|\mathbf{r}^v_i-\mathbf{r}^p_j\right|/r_p\right)$ is the
pinning energy associated with the impurities acting as pinning
centers for the vortices.
The interaction between the vortices is not significantly screened since the
penetration length is very large.
In the equations above
$\mathcal{U}_v=\left(\phi_0/4\pi\lambda\right)^2$ is the energy of
interaction between two vortices, $K_0(r)$ stands for the zero
order Hankel function, $\mathcal{U}_p<0$ is the pinning strength
created by an impurity and $f(r/r_p)$ is a rapidly decreasing
function for $r/r_p>1$. In our model we assume that the pinning
energy is much larger than the interaction between vortices
$|\mathcal{U}_p|\gg \mathcal{U}_v$ and $r_p\sim\delta$ where
$\delta$ is the lattice constant. In that case, as $N_v\ll N_p$
each vortex is preferably pinned in the close vicinity of an
impurity. As we take the London limit, which is valid for low
magnetic field and over most of the $H-T$ phase diagram in extreme
type-II superconductors such as cuprates, we assume that the size
of the vortex core is negligible and place each vortex core at
the center of a plaquette. So in the limit of strong pinning
described above, each vortex will be pinned in the center of one of
the four plaquettes surrounding a site hosting an impurity. The
plaquetes selected by the vortices are those minimizing the
interaction energy $\mathcal{E}_v$ between the vortices.

Once the impurity positions
$\left\{\mathbf{r}^p_i\right\}_{i=1,N_p}$ are fixed and the
correlated vortex positions
$\left\{\mathbf{r}^v_i\right\}_{i=1,N_v}$ are determined we are
able to calculate de quasiparticle spectrum using the BdG
equations
$\mathcal{H}(\mathbf{r})\Psi_n(\mathbf{r})=\epsilon_n\Psi_n(\mathbf{r})$,
where
$\Psi_n^\dagger(\mathbf{r})=\left(u_n^*(\mathbf{r}),v_n^*(\mathbf{r})\right)$.
It is convenient \cite{Franz1,Lages2} to perform a unitary gauge
transformation. After carying out this gauge transformation the
Hamiltonian reads
\begin{equation} \label{hamiltonian}
{\cal H} = \left( \begin{array}{cc}
\hat{h}_A & \hat{\Delta} \\
\hat{\Delta}^{\dagger} & -\hat{h}_B^{\dagger} \\
\end{array} \right)
\end{equation}
where
\[
\hat{h}_{\mu}=-t\sum_{\boldsymbol{\delta}}e^{i\mathcal{V}^{\mu}_{\boldsymbol{\delta}}(\mathbf{r})}
    \hat{s}_{\boldsymbol{\delta}}-\epsilon_F+\mathcal{U}(\mathbf{r}),
\]
\[
\hat{\Delta}=\sum_{\boldsymbol{\delta}}e^{i\mathcal{A}_{\boldsymbol{\delta}}(\mathbf{r})
+i\pi\delta_y}
\Delta(\mathbf{r,r+}\boldsymbol{\delta})\hat{s}_{\boldsymbol{\delta}}.
\]
The phase factors are given by
${\cal
V}_{\bs{\delta}}^{\mu}(\mbf{r})
=\int_{\mbf{r}}^{\mbf{r}+\bs{\delta}} \mbf{k}_s^{\mu} \cdot
d\mbf{l}$
and
$\mathcal{A}_{\boldsymbol{\delta}}(\mathbf{r})=\frac
1 2\int_{\mathbf{r}}^{\mathbf{r+}\boldsymbol{\delta}}
\left(\mathbf{k}_s^A(\mathbf{l})-\mathbf{k}_s^B(\mathbf{l})\right)\cdot
d\mathbf{l}$
where the vector
$\hbar\mathbf{k}_s^{\mu}(\mathbf{r})=m\mathbf{v}_s^{\mu}(\mathbf{r})
=\hbar\boldsymbol{\nabla}\phi^{\mu}-(e/c)\mathbf{A}(\mathbf{r})$
is the superfluid momentum vector of the effective
$\mu=A,B$-supercurrent.
This quantity can be calculated for an arbitrary configuration
of vortices like
\be
\mathbf{k}_{\mu}(\mathbf{r})=2 \pi \int \frac{d^2k}{(2 \pi)^2}
\frac{i \mathbf{k} \times \hat{z}}{ k^2+\lambda^{-2}}
\sum_i e^{i\mathbf{k} \cdot (\mathbf{r}-\mathbf{r}_i^{\mu})}.
\ee
Here $\lambda$ is the magnetic penetration length and the sum extends over all vortex
positions.
$\mbf{A}(\mbf{r})$ is the vector potential
associated with the uniform external magnetic field
$\mbf{B}=\bs{\nabla}\times\mbf{A}$.
The vortices $A$ are only visible to the
electrons and the vortices $B$ are only visible to the holes. Each
resulting $\mu$-subsystem is then in an effective magnetic field
$\mbf{B}_{\mathrm{eff}}^\mu=-\frac{mc}{e}\bs{\nabla}\times\mbf{v}^\mu_s
=\mbf{B}-\phi_0\mbf{z}\sum_i\delta(\mbf{r}-\mbf{r}_i^\mu)$
where each vortex carries now an effective quantum magnetic flux
$\phi_0$. For the case of a regular vortex
lattice\cite{Franz1,PRB}, these effective magnetic fields vanish
simultaneously on average over a unit magnetic cell containing two
vortices, one of each type. More generally, in the absence of
spatial symmetries, as it is the case for disordered systems,
these effective magnetic fields $\mbf{B}_{\mathrm{eff}}^{\mu=A,B}$
vanish if the number of vortices of the two types are in equal
number, $N_A=N_B$, and equal to the number of elementary quantum
fluxes $\phi_0$ of the external magnetic field penetrating the
system.
The sums are over nearest neighbors ($\bs{\delta}=\pm \mbf{x}, \pm
\mbf{y}$ on the square lattice) and the operator
$\hat{s}_{\bs{\delta}}$ is defined through its action on space
dependent functions, $\hat{s}_{\bs{\delta}}u(\mbf{r}) =
u(\mbf{r}+\bs{\delta})$. The energy $\mathcal{U}(\mathbf{r})$ is
the impurity potential which takes the value $U>0$ at the sites
$\left\{\mathbf{r}^p_i\right\}_{i=1,N_p}$ hosting an impurity and
zero elsewhere.
The operator
$\hat{\eta}_{\bs{\delta}}=e^{i\pi\delta_y}\hat{s}_{\bs{\delta}}$
carries the symmetry of the $d$-wave order parameter. The disorder
potential induced by the impurities and the inhomogeneous
superfluid velocities induced by the vortices strongly affect the
pairing potential
\begin{figure}
\includegraphics[width=0.8\columnwidth]{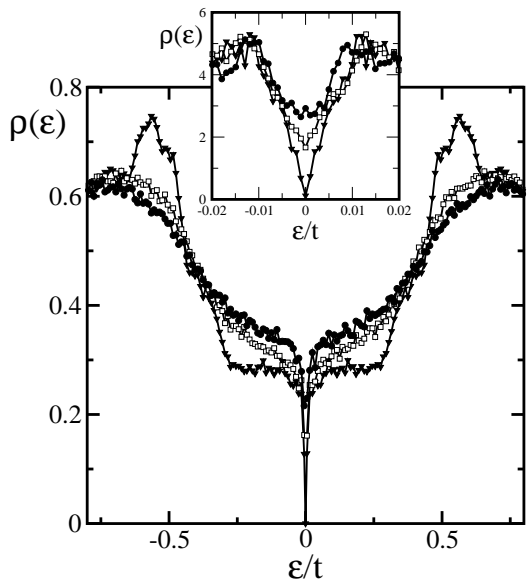}
\caption{\label{fig1} Quasiparticle density of states for a $20
\times 20$ sites system with $4\%$ of impurities, with $U=5t$,
$\epsilon_F=1.2$ and $V=-2.3t$. We consider three cases: no vortices
($\blacktriangledown$), 4 vortices pinned at locations ensuring
the minimization of the total vortex energy
($\square$), and $4$ vortices located at random ($\bullet$). For
each case the density of states is averaged over 100
configurations.}
\end{figure}
$\Delta(\mathbf{r,r+}\boldsymbol{\delta})$
defined over the link $[\mathbf{r,r+}\boldsymbol{\delta}]$ of the
two-dimensional lattice. Thus for a given configuration of the
impurity positions and of the vortex positions the pairing
potential $\Delta(\mathbf{r,r+}\boldsymbol{\delta})$ is calculated
self-consistently until convergence is obtained for each
individual link. On each lattice site we can define the amplitude
of the $d$-wave order parameter as
$\Delta_d\left(\mathbf{r}\right)=\sum_{\boldsymbol{\delta}}
(-1)^{\delta_y}\Delta(\mathbf{r,r+}\boldsymbol{\delta})$.
The amplitude of the order parameter is strongly suppressed in the
vicinity of impurities and vortices.

\begin{figure}
\includegraphics[width=0.8\columnwidth]{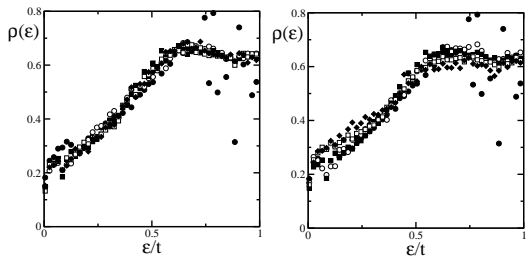}
\caption{\label{fig2} Quasiparticle density of states for a $20
\times 20$ sites. Left panel: system with $2\%$ of impurities, $4$ vortices
pinned at locations ensuring the minimization of the total vortex
energy and $U=2t (\circ), 5t (\blacksquare),
10t (\Box)$ and $100t (\blacklozenge)$. The case of 4 vortices
pinned at random in a system without impurity is also presented
($\bullet$). Right panel:
system with $4$ vortices pinned at locations
ensuring the minimization of the total vortex energy,
$U=5t$ and the following percentages of
impurity: $0\%$ [vortices pinned at random] $(\bullet), 1\%
(\circ), 2\% (\blacksquare), 4\% (\Box)$ and $6\%
(\blacklozenge)$.}
\end{figure}

As the effective magnetic fields experienced by the
particles and the holes vanish on average, within the gauge transformation
we are allowed to use periodic boundary conditions
on the square lattice ($\Psi(x+nL,y+mL)=\Psi(x,y)$ with $n,m\in
\Bbb{Z}$). The $L\times L$ original lattice becomes then a
magnetic supercell where the impurities are placed at random and
where the vortices are placed in such a way as to minimize their
total energy. The disorder induced by the
impurities in the system is then established over a length $L$.
Thus in order to compute the eigenvalues and eigenvectors of the
Hamiltonian (\ref{hamiltonian}) we seek for eigensolutions in the
Bloch form
$\Psi^\dagger_{n\mbf{k}}(\mbf{r})=e^{-i\mbf{k}\cdot\mbf{r}}
(U^*_{n\mbf{k}},V^*_{n\mbf{k}})$ where $\mbf{k}$ is a point of the
Brillouin zone. We diagonalize then the Hamiltonian
$e^{-i\mbf{k}\cdot\mbf{r}}\mathcal{H}e^{i\mbf{k}\cdot\mbf{r}}$ for
a large number of points $\mbf{k}$ in the Brillouin zone and for
many different realizations (around 100) of the random impurity
positions and of the correlated vortex positions.

\begin{figure}
\includegraphics[width=0.9\columnwidth]{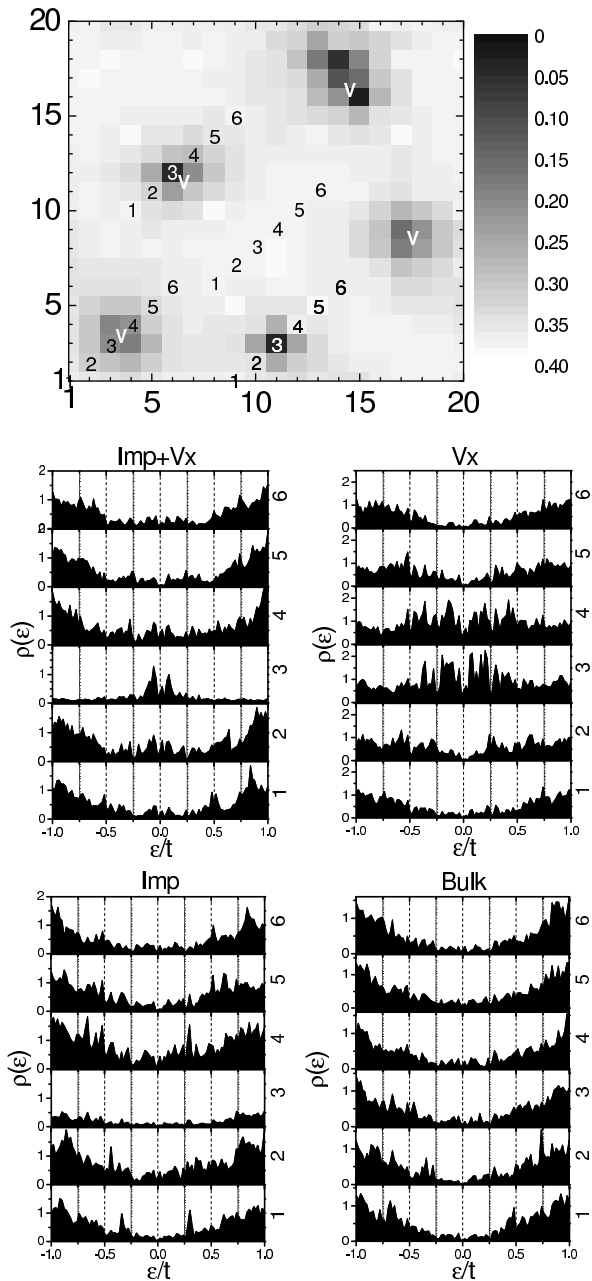}
\caption{\label{fig3} Quasiparticle local density of states for
different points on the lattice. The impurity concentration is
$0.75\%$. We have chosen a particular distribution of the
impurities and vortices where we can find the various
possibilities of a site far from any impurity or vortex
(\textbf{\textsf{Bulk}}), an impurity site with no vortex nearby
(\textbf{\textsf{Imp}}), a site close to a vortex and no impurity
nearby (\textbf{\textsf{Vx}}) and an impurity site with a vortex
attached (\textbf{\textsf{Imp$+$Vx}}). For each of these 4
possibilities we have calculated the quasiparticle local density
of states along a path of six numbered sites $1,\ldots,6$. On the
top most pannel the 4 particular paths are shown on the top of the
corresponding profile of the $d$-wave order parameter
$\Delta_d(\mbf r)$. The path located on the center of the lattice
corresponds to the \textbf{\textsf{Bulk}} case, the one centered
on the site (6,12) corresponds to the \textbf{\textsf{Imp$+$Vx}}
case, the one centered on the site (3,3) corresponds to the
\textbf{\textsf{Vx}} case, and the one centered on the site (11,3)
correspond to the \textbf{\textsf{Imp}} case.}
\end{figure}

In Fig. \ref{fig1} we show the density of states averaged over
$100$ configurations for a moderate impurity concentration and for
a typical value of $U=5t$. We compare the cases with no vortices
and with a low vortex density (considering both an energetically
favorable distribution of the vortices and a fully random
configuration). The self-consistent calculation of the order
parameter gives a maximum amplitude, $\Delta_0$, of the order of
$0.5t$. It is clear that when there are no vortices present,
coherence peaks appear at $\epsilon \sim \Delta_0$. If the
vortices penetrate the sample these peaks disappear.
Without vortices the density of states vanishes at
zero energy as found before and if we include the magnetic field
the density of states becomes finite \cite{Lages2}. Also it is
clear that if the vortices are fully randomly distributed the DOS
is larger than the one where the vortices tend to be pinned at the
impurity sites. The results are therefore qualitatively similar to
the ones obtained when there are no impurities present ($U=0$)
\cite{Lages2}. In Fig. \ref{fig2}, we show the
influence of the impurity concentration and of the repulsive local
potential, $U$. Changing the impurity concentration leads to no
qualitative difference except that there is a slight increase in
the DOS. The same happens with the change of $U$. Both sets of
results indicate that the strong effect is the scattering off the
vortices.

More detailed information about the scattering of the
quasiparticles is obtained calculating the LDOS for a given
impurity/vortex configuration. It is defined by
\[
\rho( \vec{r},\epsilon) = \sum_n \left( |u_n(\vec{r})|^2 +
|v_n(\vec{r})|^2 \right) \delta (\epsilon -\epsilon_n).
\]
In Fig. \ref{fig3} we compare the LDOS at four different sites.
i) At a site in the bulk the band-structure profiles are somewhat similar
to the case of a vortex lattice \cite{Lages2}. The coherence peaks
are evident, the zero energy density of states is very small (but
not strictly zero), and the low energy peaks are smeared out. ii)
At an impurity site (and no vortex nearby) the same structure is
apparent except that since the impurity potential is repulsive the
density of states at the impurity site is considerably depleted at
low energy (for instance in the very large $U$ limit the density
of states is virtually zero at the impurity site).
iii) In the vicinity
of a vortex site (but far from any impurity) the structure is very
similar to the case obtained before \cite{Lages2} with no
coherence peaks close to the vortex and an enhanced zero energy
density of states near the vortex. iv) Finally the interesting
case of a location where a vortex is bound to an impurity reveals
that the coherence peaks are recovered very close to the vortex.
Also, the low energy density of states at the impurity site is
increased with respect to the (\textbf{\textsf{Imp}}) case, due to the vortex nearby. However
the density of states is considerably smaller than for the case (\textbf{\textsf{Vx}}) of a vortex far from any impurity.

The results obtained previously for the vortex lattice case with no
impurities explain qualitatively the DOS results but are not realistic.
In the experimental systems disorder is present and its effect must be taken into
account.
The results show that the dominant effect on the quasiparticle DOS
is due to the vortex scattering. The presence
of an impurity basically renormalizes the DOS except when the impurity is strongly repulsive
where the density of states is significantly depressed near the impurity. This is the unitary
limit where a gapped system is predicted in the absence of magnetic field \cite{Altland}.
The quantum effect originated in the Aharonov-Bohm scattering of the quasiparticles
circulating around a vortex line has been shown to have considerable effects on the
density of states \cite{PRB,Melnikov}. Significant changes with respect to the
classical Doppler shift effect \cite{Volovik} occur at low energies \cite{PRB}.
The results obtained in this paper show that the addition of impurity scattering is
not very significant and the Berry phase is dominant, as argued before \cite{Franz1,Lages2}.
The results are very similar to the ones obtained experimentally with STM in ref.
\cite{Pan}, except for the finite density of states at zero energy in the vicinity
of an isolated vortex. The experimental results are therefore more consistent
with the situation where all or most of the vortices are pinned to the impurity sites.
At these points, even though there is an enhancement of the low energy density of states
with respect to an impurity with no vortex attached, the increase is reduced by the
presence of the impurity with respect to the case of a vortex but no impurity nearby.


\begin{thebibliography}{99}
\bibitem{dwave} D. J. van Harlingen, Rev. Mod. Phys. {\bf 67}, 515 (1995).
\bibitem{Hoffman1} J. E. Hoffman, E. W. Hudson, K. M. Lang, V. Madhavan, H. Eisaki, S. Uchida, and J. C. Davis, Science {\bf 295}, 466 (2002).
\bibitem{Pan} S. H. Pan, E. W. Hudson, A. K. Gupta, K.-W. Ng, H. Eisaki, S. Uchida, and J. C. Davis, Phys. Rev. Lett. {\bf 85}, 1536 (2000).
\bibitem{Franz1} M. Franz and Z. Te\v{s}anovi\'c, Phys. Rev. Lett. {\bf 80}, 4763 (1998).
\bibitem{magcdw} S.C. Zhang, Science {\bf 275}, 1089 (1997); D.P. Arovas, A. J. Berlinsky, C. Kallin, and Shou-Cheng Zhang, Phys. Rev. Lett. {\bf 79}, 2871 (1997); H-Yi Chen and C.S. Ting, cond-mat/0306232;
cond-mat/0402141.
\bibitem{Lages2} J. Lages, P. D. Sacramento and Z. Te\v{s}anovi\'c,
        Phys. Rev. B {\bf 69}, 094503 (2004).
\bibitem{Altland} A. Altland, B. D. Simons and M. R. Zirnbauer,
                 Phys. Rep. {\bf 359}, 283 (2002).
\bibitem{Atkinson2} W. A. Atkinson, P. J. Hirschfeld, and A. H. MacDonald, Phys. Rev. Lett.
{\bf 85}, 3922 (2000).
\bibitem{Bloch} M. Franz and Z. Te\v{s}anovi\'c,
             Phys. Rev. Lett. {\bf 84}, 554 (2000).
\bibitem{Gorkov} L. P. Gorkov and J. R. Schrieffer,
                 Phys. Rev. Lett. {\bf 80}, 3360 (1998).
\bibitem{Volovik} G. E. Volovik,
                  Pis'ma Zh. \'Eksp. Teor. Fiz. {\bf 58}, 457 (1993);
                  [JETP Lett. {\bf 58}, 469 (1993)].
\bibitem{K.} J. Ye, Phys. Rev. Lett. {\bf 86}, 316 (2001); D. V. Khveshchenko and A. G.
Yashenkin, Phys. Rev. B {\bf 67}, 052502 (2003).
\bibitem{aprile} I. Maggio-Aprile, Ch. Renner, A. Erb, E. Walker, and \O. Fischer, Phys. Rev. Lett. {\bf 75}, 2754 (1995).
\bibitem{Atkinson} W. A. Atkinson, P. J. Hirschfeld, A. H. MacDonald, and K. Ziegler, Phys. Rev. Lett. {\bf 85}, 3926 (2000).
\bibitem{pinning} G. Blatter, M. V. Feigel'man, V. B. Geshkenbein, A. I. Larkin, and V. M. Vinokur, Rev. Mod. Phys. {\bf 66}, 1125 (1994).
\bibitem{PRB} O. Vafek, A. Melikyan, M. Franz, and Z. Te\v{s}anovi\'c, Phys. Rev. B {\bf 63}, 134509 (2001);
\bibitem{Melnikov} A.S. Mel'nikov, Phys. Rev. Lett. {\bf 86}, 4108 (2001).
\end{thebibliography}
\end{document}